\documentclass{article}

\usepackage{PRIMEarxiv}

\usepackage[utf8]{inputenc} %
\usepackage[T1]{fontenc}    %
\usepackage{amssymb}
\usepackage{amsmath}
\usepackage{amsfonts}
\usepackage{booktabs}       %
\usepackage{multirow}
\usepackage{array}
\usepackage{capt-of}      %
\usepackage{nicefrac}
\usepackage{microtype}      %
\usepackage{graphicx}
\graphicspath{{figures/}}
\usepackage{fancyhdr}       %
\usepackage[numbers,sort&compress]{natbib}
\usepackage{url}
\usepackage{hyperref}       %
\hypersetup{colorlinks=true, linkcolor=blue, citecolor=blue, urlcolor=blue}

\title{Endorsement Without New Evidence:\\
How Sequential Voting Inflates Mandates\\
in Online Community Governance
\thanks{This manuscript is currently under peer review.}}

\author{
  Zihan Chen \\
  Stevens Institute of Technology \\
  Hoboken, NJ, USA \\
  \texttt{zchen61@stevens.edu} \\
  \And
  Lei Nico Zheng \\
  University of Massachusetts Boston \\
  Boston, MA, USA \\
  \texttt{lei.zheng@umb.edu} \\
  \And
  Di Zhu \\
  Stevens Institute of Technology \\
  Hoboken, NJ, USA \\
  \texttt{dzhu1@stevens.edu} \\
}

\begin{document}
\maketitle
\setcounter{footnote}{0}

\begin{abstract}
Online communities often treat large support margins in public elections as strong mandates. We argue that such margins can overstate the independent scrutiny behind a decision. Using 198{,}275 free-text rationales from Wikipedia admin elections, we introduce vote--text divergence, a measure that flags a decisive vote paired with a thin, deferential rationale. Divergence rises as voters arrive later, even after controlling for voter and election fixed effects. The pattern is consistent with information saturation: once prior text is accounted for, arrival order no longer predicts divergence, while accumulated prior evidence does. The effect is strongest among peripheral voters in the co-voting network. Yet divergence does not predict worse post-promotion outcomes, such as administrative activity or survival. Public tallies can therefore weaken the scrutiny signal even while selecting capable administrators: a margin may appear to reflect more consensus and support than it actually contains.
\end{abstract}

\keywords{online community governance \and online voting \and
observational learning \and information saturation \and bounded rationality}

\section{Introduction}
Large online communities rely on voting and decentralized self-organization to allocate authority. Wikipedia, open-source projects, Q\&A sites, and major forums often promote ordinary members into administrative roles through some form of peer vote, with the resulting tally read as the community's verdict \cite{forte2009decentralization,kraut2012building,johnson2015emergence}. The common assumption is that a larger support margin signals broad community consensus and a stronger mandate.

A tally, however, counts votes, and votes can carry unequal information. Consider a voter who arrives once an election already shows ninety-five percent support across eighty votes. She can read the candidate's contributions and report what she independently finds, or she can defer, writing ``per nom'' and moving on. Deferring is cheap, and at that point her single vote will not change the outcome \cite{bikhchandani1992theory,feddersen1996swing}. When many late voters defer, the count keeps rising while the independent scrutiny behind it does not.

Why does this matter if the community still selects capable leaders or administrators? Because the margin is rarely confined to the original election; it is also read as a scrutiny signal: evidence of how much independent scrutiny and support the community invested in the candidate. For example, bureaucrats may treat near-unanimous support as stronger legitimacy when closing a discussion \cite{forte2009decentralization,leskovec2010governance}, and later community members may read a past support ratio as part of a candidate's reputation \cite{leskovec2010signed,west2014exploiting}. If a ninety-five percent margin reflects less independent scrutiny than its size suggests, these later uses can inherit the same overstatement. 

We make this gap observable through the text beside each vote. When a voter writes ``Support, per above,'' she casts the same $+1$ as one who cites specific edits and policy pages, yet only the second adds information to the record. We define vote--text divergence as a decisive vote paired with a thin, deferential rationale that adds little evidence. Prior work on online influence infers conformity from actions or ratings alone \cite{muchnik2013social,lee2015follow}; because adminship votes record the reason beside the act, we can instead observe directly when a vote's sign and the information behind it come apart.

This setup leads to three questions. First, does a voter's position in an election shape how much independent scrutiny their vote adds to the record? Second, what mechanism drives any such pattern: conformity to the visible majority, or saturation of the evidence already written? Third, does this deference change who the community promotes, or only how much its margins should be trusted?

To answer these questions, we study the Stanford SNAP wiki-RfA corpus \cite{west2014exploiting}. Requests for Adminship (RfA) is Wikipedia's process for granting administrator rights: editors cast public, signed votes with short written rationales over about a week, and a bureaucrat review the community discussion and grant adminship to the candidate if reach consensus. The corpus contains 198{,}275 signed votes with rationales over 4{,}003 elections from 2003 to 2013, joined to the complete public record of each elected admin's later administrative actions. The observation period ends before the wide spread of Wikipedia bots and large language models \cite{zheng2019roles}, which offers a clean record of human language use in online community governance under a simple, visible feedback architecture.

The empirical pattern is clear and internally consistent. First, we found that vote--text divergence rises with arrival position, and later votes carry more deference language and less evidence. Second, the evidence is not consistent with a simple visible majority explanation. Divergence does not increase as the tally becomes more decisive, and the gradient is actually steeper in contested elections than in landslides. What it tracks instead is the accumulation of information: once we control for how much text has already been written, the arrival-position effect disappears and the volume of prior text absorbs it. This is consistent with information saturation under bounded rationality: as the thread grows, a later voter sees that the evidentiary space is mostly covered and responds by restating agreement rather than adding new scrutiny. Third, the behavior is concentrated among network-peripheral voters, who are plausibly both less pivotal and less invested in community governance.

A natural fear is that this behavior corrupts the community's choices. It does not, at least not in anything we can measure. Election-level divergence is nearly orthogonal to the winning margin and predicts no deterioration in any post-promotion outcome, from action volume to removal for cause, and the marginal admit near the promotion threshold is no worse on these measures. Tellingly, the bureaucrats who exercise discretion in close cases do not discount textually divergent elections, so the institution does not correct the signal on its own. The damage therefore lands on the signal the margin sends, while the selection it informs holds up: the community still promotes capable administrators, yet the margin overstates the scrutiny behind them, so the burden of correction falls on how the record is displayed and aggregated.

This paper makes three contributions. First, we introduce vote--text divergence, a reproducible, text-based measure that quantifies, from short vote rationales, when a decisive vote is paired with a thin, deferential reason. Second, we show that this divergence is ordered by a voter's arrival position and is best explained by information saturation rather than by the decisiveness of the tally. Third, we connect the behavior to governance outcomes: it concentrates among network-peripheral voters, yet predicts no deterioration in any observable post-promotion outcome, so a large margin can misstate the scrutiny behind it even when the selection itself looks unharmed.

\section{Related Work and Theory}

\subsection{Governance by visible voting in online communities}

Online communities institutionalize authority through peer evaluation \cite{forte2009decentralization,kraut2012building,ren2012building}, and adminship can be read as a formalized case of leadership emergence, in which authority arises from community interaction rather than external appointment \cite{johnson2015emergence}. This is one instance of governance by aggregation, in which authority comes from many distributed evaluations collapsed into a single visible verdict rather than from a central decision-maker, a mechanism IS research increasingly studies for both human and AI-driven systems \cite{berente2021managing}. Wikipedia's Requests for Adminship (RfA) is the canonical setting. It has been studied as a promotion process \cite{leskovec2010governance,burke2008mopping} and as a signed, sequential social network \cite{leskovec2010signed,west2014exploiting}. Burke and Kraut \cite{burke2008mopping} model who is promoted, and Leskovec et al. \cite{leskovec2010governance} model the deliberative process. We shift attention from who wins to what the voting record means, and ask whether a larger margin reflects more independent scrutiny or more deferential repetition.

\subsection{Observational learning, saturation, and pivotality}

When agents act in sequence and observe prior actions, private signals can stop entering the public record once a direction is visible, and later actors imitate. This is the logic of observational learning and, in its strong form, the informational cascade \cite{bikhchandani1992theory}. We treat the cascade as an alternative account, distinct from the bandwagon and information saturation forces we define below, and test it directly in Section~5.2. Displaying a prior aggregate is itself enough to move behavior. A single up-vote raises final ratings \cite{muchnik2013social}, visible popularity reorders cultural markets \cite{salganik2006experimental}, and exposing historical behavior is a platform-design choice that can trigger cascades \cite{duan2009informational}. In IS and marketing, observational learning and peer influence shape adoption and ratings as people weigh the crowd against their own signals \cite{zhang2010sound,lee2015follow,moe2012online,bapna2015online}. This stream asks whether visible prior behavior changes later actions. We ask a different question: whether a later voter's vote still adds evidence to the record, given that the same support vote can carry genuine scrutiny or mere deference.

Two forces can produce deference in this setting, and they carry different implications. The first is a bandwagon force driven by visibility: a salient running count lowers the perceived need to add independent scrutiny, and it makes a late voter unlikely to be pivotal, which weakens the incentive to invest in a costly private signal \cite{feddersen1996swing,sunstein2006infotopia}. This is the swing voter's logic: when one vote will not change the outcome, the return to private scrutiny falls. The second is an information saturation force: processing an expanding discussion record is costly, so a boundedly rational voter who judges that the relevant evidence has largely been stated and responds by restating agreement \cite{simon1955behavioral,marchsimon1958organizations}. Saturation is the within-thread analogue of information overload: in online interaction spaces, as the message record accumulates, participants shift toward simpler, lower-effort responses rather than adding new content \cite{jones2004overload,eppler2004overload}, so the marginal value of adding to the record falls as the record grows.

This view also recasts a distinction the record blurs. Empty deference substitutes the crowd's judgment for the voter's own private scrutiny. Efficient deference avoids restating evidence that earlier voters have already supplied. We treat efficient deference as one channel through which independent scrutiny becomes concentrated in early votes. The key point holds for both kinds: even when deference is efficient rather than empty, the vote margin stops counting independent evidentiary contributions one for one. Because our construct measures what enters the public record, it captures this visible deficit in either case, and we are explicit about what it cannot separate.

\subsection{Who defers: experience and motivation}

Our peripheral-voter result connects to two literatures. Marketing research on herding in online ratings finds that conformity to a visible consensus depends on evaluator experience, with less experienced raters more prone to follow the crowd \cite{sunder2019herding}. Work on contribution motives distinguishes intrinsic from image-related utility \cite{toubia2013intrinsic}. Core members invested in the community's health have reason to supply independent scrutiny, while peripheral participants may gain more from signaling alignment than from costly scrutiny. Both literatures predict that deference should fall disproportionately to peripheral voters, which is what we find.

\subsection{Hidden profiles and text as data}

Group-decision research on the hidden profile shows that information shared among members crowds out the unique signals only a few hold, so groups fail to surface what only a minority knows \cite{stasser1985pooling,dennis1996information}. Vote--text divergence makes that failure observable at scale. Measuring it requires reading short, conventional text, where the central threat to validity is that language reflects convention or politeness rather than judgment \cite{brown1987politeness,godes2004using,humphreys2018automated,berger2020uniting}. The measurement task is therefore to recover an invariant quantity, the substance a voter actually contributes, when the most salient surface feature, the wording and length of a short comment, is also the one that varies most with convention. That identification problem is not specific to text; it recurs wherever a dominant observable attribute is confounded with the latent one of interest, as in cloth-changing person re-identification, where the most salient visual cue must be discounted to recover identity-invariant structure \cite{ding2026synergistic}. We treat the threat as central to construct validity, model deference language as its own dimension, and test the construct against alternatives in Section~5.5.

\subsection{Hypotheses}

We fix the vocabulary before stating hypotheses. Table~\ref{tab:glossary} locks each core construct to a single canonical term and meaning, so that the same idea is named the same way throughout.

\begin{table}[thb]
\centering
\caption{Construct glossary: canonical term and meaning in this paper.}
\label{tab:glossary}
\small
\begin{tabular}{@{}p{3.4cm}p{12.4cm}@{}}
\toprule
Term & Meaning in this paper \\
\midrule
Vote--text divergence & A decisive vote paired with a thin, deferential rationale; the negative residual of informative conviction on vote sign, decisiveness, and length (\S3.2). Measures scrutiny entered into the public record. \\
Informative conviction & The independent substance a single rationale carries, summed over five text dimensions (\S3.2); divergence is its residualized negative. \\
Information saturation & A late voter's choice to restate agreement once the evidentiary space looks exhausted; our mechanism. Distinct from fatigue or information overload. \\
Deferring & Casting a decisive vote with a low-information rationale (e.g., ``per nom''). Deference language is the compliance text dimension; empty vs efficient deference is the theoretical split. \\
Independent scrutiny & A voter's own evaluation of the candidate; we measure only the part entered into the public record. \\
Network-peripheral voter & A voter with low co-voting eigenvector centrality (\S4). \\
\bottomrule
\end{tabular}
\end{table}

We frame RfA as observational learning under a public aggregate. Early voters report private signals; the running count and the discussion become public; and later voters can defer to them. Vote--text divergence is the wedge between a decisive vote and a thin reason. We state the gradient as our reading of the evidence and test competing accounts in Section~5.5.

\begin{itemize}
\setlength\itemsep{0.1em}
\item \textbf{H1a (Deference gradient).} Within an election, vote--text divergence rises with arrival position. Later votes carry more deference language and less evidentiary grounding.
\item \textbf{H1b (Saturation mechanism).} The gradient is mediated by the cumulative prior discussion. Once the volume of prior text is held fixed, arrival position no longer predicts divergence.
\item \textbf{H2 (Peripheral locus).} The gradient is concentrated among network-peripheral voters, those with low co-voting eigenvector centrality (operationalized in \S4). Central, experienced insiders defer less.
\item \textbf{Consequences (tested empirically).} If deference corrupts the decision itself, beyond its appearance, election-level divergence should predict weaker downstream outcomes. We test this directly.
\end{itemize}

\section{Data and Measurement}

\subsection{The wiki-RfA corpus}

We use the Stanford SNAP wiki-RfA dataset \cite{west2014exploiting}. Each record is a signed vote ($+1$ support, $0$ neutral, $-1$ oppose) by one editor on one candidate, with the year, a timestamp, and a free-text rationale in wiki markup. Parsing yields 198{,}275 votes (144{,}451 support, 41{,}176 oppose, 12{,}648 neutral) by voters on 3{,}497 candidates. Clustering a candidate's votes into runs separated by gaps over thirty days gives 4{,}003 elections, of which 1{,}903 promoted the candidate. Text is present for 96.4\% of votes but is short, with a median of 14 tokens, so we design the measure around effort and substance rather than long-form sentiment. Timestamps parse for 91.7\% of votes, which is enough to order voters within an election.

\subsection{Vote--text divergence}

We score each rationale on five dimensions with a transparent rule-and-lexicon (dictionary-based) text-as-data system applied to the raw markup, following the validity standards for automated text analysis in \cite{humphreys2018automated,berger2020uniting}. A dictionary method is the appropriate choice here because the rationales are short and conventional, where supervised classifiers have little signal to learn from, and because a transparent lexicon makes each dimension auditable and reproducible, which \citet{humphreys2018automated} identify as a core validity requirement. For example, a one-line ``Support per nom.'' offers almost no features for a trained classifier but is unambiguously scored by a deference lexicon. Valence is sentiment, entered as intensity ($\lvert z\rvert$) so that a substantive oppose is not penalized for negative sentiment. Certainty is confidence markers net of hedges. Specificity counts concrete particulars, such as numbers, named Wikipedia processes, and article links. Grounding counts evidentiary references, policy and guideline shortcuts ([[WP:\dots]]), diffs, and permalinks, read off the raw text before markup is stripped. Compliance captures deference language, such as ``per nom,'' ``per above,'' bare agreement, and one-line votes.

These five dimensions combine into informative conviction, a single index of how much independent substance a rationale carries: high when the text is confident, specific, and grounded in evidence, and low when it leans on deference language. We standardize each dimension and define informative conviction as
\[
\begin{aligned}
&\lvert z(\text{valence})\rvert + z(\text{certainty}) + z(\text{specificity}) \\
&\qquad {}+\, z(\text{grounding}) - z(\text{compliance}).
\end{aligned}
\]
Vote--text divergence is then the negative residual of informative conviction regressed on the vote sign, its decisiveness, and log length. A high value is a decisive vote whose text carries less substance than its sign and length predict. Because length is partialled out, divergence is not a relabeling of brevity; its correlation with token count is only $-0.16$. The construct measures only the independent scrutiny entered into the public record, setting aside whatever a voter may have judged privately. That is the right object for our claim, which concerns what a margin visibly represents.

The construct separates four things that short rationales blur, summarized in Table~\ref{tab:typology}: brevity, deference to prior voters, the presence of evidence, and candidate-specific scrutiny. High divergence is a decisive vote that is deferential and evidence-free. Low divergence names specific behavior or policy, whether support or oppose, and need not be long. We include the hard cases on purpose. A short vote that points to a specific prior argument may be efficient deference rather than absent scrutiny, and a generic ``no concerns after review'' asserts scrutiny without showing it. These ambiguous cases are what a human-coded validation must adjudicate.

\begin{table}[thb]
\centering
\caption{Construct typology with representative rationales (lightly cleaned; usernames removed).}
\label{tab:typology}
\small
\begin{tabular}{@{}p{7.0cm}ccc@{}}
\toprule
Rationale & Defers? & Evid.? & Div. \\
\midrule
``Support, per above.'' & Yes & No & High \\
``Support per nom.'' & Yes & No & High \\
``Support; 5 yrs, AfD/CSD backlog work.'' & No & Yes & Low \\
``Strong oppose: content creation $\approx$ 0.'' & No & Yes & Low \\
``Per X's diff evidence above; support.'' & Yes & Indirect & Ambig. \\
``Support, no concerns after review.'' & Weak & No & Ambig. \\
\bottomrule
\end{tabular}
\end{table}

Two checks support the construct. The most-divergent decile of votes scores far above average on deference language ($z=+1.49$) and below on grounding ($z=-0.12$). The least-divergent decile is the reverse, with deference at $z=-0.39$ and grounding at $z=+1.00$. The two deciles have similar length, 42 versus 57 tokens, so the contrast is about substance rather than brevity. Every result below also replicates under a principal-component composite of the five dimensions and under dropping the valence dimension entirely. The second check matters because it shows the gradient is driven by a decline in structural substance, namely evidence and specificity, while affective intensity stays flat. We have not yet conducted human gold-standard coding, which is the single most valuable next step. We release the scoring code and a stratified 1{,}387-comment instrument so that a coded validation, which can separate efficient from empty deference, can be run directly.

\subsection{Observable governance outcomes}

For each elected admin we reconstruct objective, pre-specified post-promotion outcomes from public Wikimedia logs (the logevents and rights APIs), joined by username, with a 100\% match rate. We build an index of administrative activity and survival, the mean of standardized components: log action volume, action-type breadth, tenure length, sustained activity, and not having been removed for cause, where for-cause removals are separated from benign resignations and inactivity. These are observable governance outcomes rather than a validated measure of quality, a limitation we return to. All reporting is aggregate, and no individual admin is labeled.

\section{Empirical Strategy}

\textbf{Deference gradient (H1a).} For votes in elections with at least ten votes, we regress per-vote divergence on the voter's standardized arrival position, where 0 is first and 1 is last. We use three increasingly demanding specifications: election fixed effects, voter fixed effects, and two-way voter-and-election fixed effects estimated by alternating projection. The two-way model compares later and earlier votes after absorbing both stable voter-level writing tendencies and election-level conditions. Identification comes from voters who appear across many elections and elections with many voters. We add a within-election permutation test on arrival order, decompose the gradient into its deference and grounding components, and replicate under the alternative composite.

\textbf{Saturation versus bandwagon (H1b).} To ask whether the gradient is bandwagon-driven, we add the tally state the voter sees at arrival, namely the support share and the cushion above the promotion threshold. To test the saturation account, we add the cumulative prior text, and then split that prior text into prior evidence and prior deference language to ask which form of accumulation drives later deference.

\textbf{Peripheral locus (H2).} We build the co-voting network, in which two editors are linked when they vote in the same election, and compute each voter's eigenvector centrality from the voter-by-election incidence matrix. We interact centrality with arrival position under election fixed effects, and we repeat the test with a strictly lagged measure of embeddedness, the count of prior elections a voter has joined. Embeddedness is a robustness proxy for centrality that serves only to rule out look-ahead in the static centrality measure.

\textbf{Consequences.} Among elected admins we regress each governance outcome on election-mean divergence, with the margin, size, and year fixed effects. We add a fuzzy regression discontinuity that instruments being elected with crossing the discretionary 75\% promotion threshold. We also test whether bureaucrats, who exercise discretion in the grey zone, implicitly discount divergent elections.

\section{Results}

\subsection{Deference rises with arrival order (H1a)}

\begin{figure}[tbp]
\centering
\begin{minipage}[t]{0.45\textwidth}\vspace{0pt}
  \centering
  \captionof{table}{H1: arrival order raises divergence, under increasingly demanding fixed effects (SE clustered by election; 194{,}609 votes).}
  \label{tab:h1}
  \vspace{4pt}
  \small
  \begin{tabular}{@{}lcc@{}}
  \toprule
  Specification & Position coef & $p$ \\
  \midrule
  Election FE                  & $+0.328$ & $<0.001$ \\
  Voter FE                     & $+0.438$ & $<0.001$ \\
  Two-way (voter+elec.) FE     & $+0.448$ & $<0.001$ \\
  \midrule
  \multicolumn{3}{@{}l}{\footnotesize Decomposition (election FE):} \\
  \;\; Deference language ($z$)    & $+0.119$ & $<0.001$ \\
  \;\; Evidentiary grounding ($z$) & $-0.080$ & $<0.001$ \\
  \;\; Permutation $p$             & \multicolumn{2}{c}{$<0.005$} \\
  \bottomrule
  \end{tabular}
\end{minipage}\hfill
\begin{minipage}[t]{0.52\textwidth}\vspace{0pt}
  \centering
  \includegraphics[width=\linewidth]{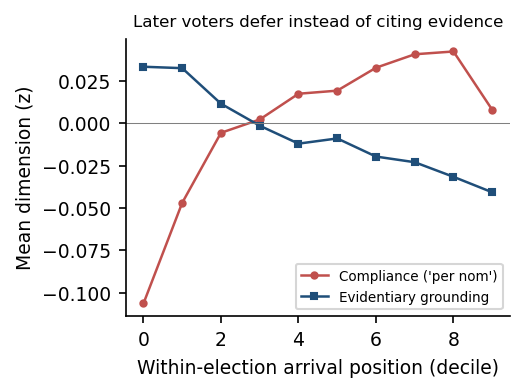}
  \caption{Within an election, later votes carry more deference language and cite less evidence.}
  \label{fig:decomp}
\end{minipage}
\end{figure}

Table~\ref{tab:h1} reports the gradient on 194{,}609 votes across 3{,}339 elections. Under election fixed effects, moving from the first to the last voter raises divergence by $0.33$ (SE $0.018$, $p<0.001$). The estimate is, if anything, stronger under tighter identification, at $0.44$ with voter fixed effects (10{,}342 distinct voters) and $0.45$ with two-way voter-and-election fixed effects ($p$ below $10^{-100}$). Because the two-way model absorbs stable voter-level and election-level composition, the gradient does not reflect who tends to arrive late or which elections run long. A permutation test that reshuffles arrival order within each election never reproduces the observed slope in 200 draws ($p<0.005$), and the gradient replicates under the principal-component composite. Decomposing the same variation, later arrival raises deference language ($+0.119$, $p<0.001$) and lowers evidentiary grounding ($-0.080$, $p<0.001$) and specificity ($-0.061$, $p<0.001$). Figure~\ref{fig:decomp} shows the two curves crossing as an election proceeds.

\subsection{Information saturation, not consensus (H1b)}

The gradient is not driven by the visible majority. Adding the tally state at arrival, the effect does not strengthen as the count becomes more decisive. The cushion of support above the promotion threshold enters negatively ($-1.33$, $p<0.001$), so votes cast when the outcome looks more settled are, if anything, less divergent. Splitting elections by their final margin, the gradient is steeper in contested races, with a position coefficient of $0.45$, than in landslides, where it is $0.25$. This speaks against the count-driven forms of conformity, both the visible-majority bandwagon and the classic informational cascade, which would each sharpen as the tally becomes more decisive. It does not, however, separate saturation from conformity to the accumulated evidence itself, which makes a similar prediction; we return to that limit in Section~5.5.

What the gradient tracks instead is how much has already been said. We measure the cumulative prior discussion as the total number of rationale tokens written by earlier voters in the same election, before the focal vote, and enter it in logs with standard errors clustered by election. It is correlated with arrival position, as expected, at $0.63$. When we add it, the arrival-position effect falls from $0.33$ to a precise zero ($+0.02$, $p=0.54$), while prior text enters strongly ($+0.09$, $p<0.001$) and absorbs it. Later voters defer because of how much discussion precedes them; once that is held fixed, their position in line carries no further weight.

We can say more about what kind of accumulation matters. We split the prior text into prior evidence, the evidentiary grounding already entered, and prior deference language, the compliance language already entered, and include both. Prior evidence drives later divergence ($+0.34$, $p<0.001$), while prior deference language does not, in fact entering negatively once evidence is held fixed ($-0.19$, $p<0.001$). What draws later deference is the evidence already on the record, while accumulated prior deference plays no such role. This evidence-driven pattern is the evidentiary form of information saturation rather than social mimicry of a deferential norm. The same split separates saturation from simple fatigue. If the decline reflected fatigue from reading a long thread, contested and landslide threads of equal length should decline in parallel; instead the gradient is steeper in contested races, which points to a deliberate processing choice once the evidentiary space looks exhausted rather than exhaustion alone.

\subsection{Deference is a peripheral-voter act (H2)}

We build the co-voting network and measure each voter's eigenvector centrality. Interacted with arrival position under election fixed effects, centrality flattens the gradient. The position-by-centrality coefficient is $-0.16$ ($p<0.001$), so the gradient is much steeper for network-peripheral voters than for central insiders. Two robustness checks address the obvious concerns. First, the result holds with a simple activity proxy, which is expected because activity and centrality are highly correlated. Second, and more important, the static network centrality could in principle use information from later elections. We therefore recompute the same construct as a strictly lagged embeddedness proxy, the number of prior elections in which the voter participated, using only votes cast before the focal one; embeddedness here stands in for centrality rather than naming a separate construct. The interaction with arrival position remains negative and significant ($-0.08$, $p<0.001$), so the locus result does not depend on look-ahead. For example, late in a thread a central, experienced editor is the one still apt to write a substantive ``Support; 5 yrs, AfD/CSD backlog work,'' while a peripheral voter arriving at the same point is the one who writes ``Support, per above.'' The pattern fits evidence that conformity rises with evaluator inexperience \cite{sunder2019herding} and that peripheral participants gain more from signaling alignment than from costly scrutiny \cite{toubia2013intrinsic}. We read peripheral status as a proxy for weaker incentives to add costly independent scrutiny, stopping short of direct proof of non-pivotality.

\subsection{No detectable deterioration in observable post-promotion outcomes}

Divergence is not the headline tally in disguise. Among elected admins, election-mean divergence correlates with the support fraction at only $-0.12$. The decisive question is whether the behavior changes who governs, and it does not. Table~\ref{tab:cons} regresses governance outcomes on standardized election-mean divergence, with the margin, size, and year fixed effects ($N=1{,}902$ elected admins, all matched to their log record). The activity-and-survival index shows no relationship ($-0.011$, $p=0.51$), and it is stable under the alternative composite and across leave-one-year-out samples. The volume, breadth, and for-cause-removal components are individually null. The only significant coefficients point the opposite way from the corruption hypothesis, with higher-divergence admins serving modestly longer. We read these as descriptive, since election-level divergence also proxies candidate popularity, and the within-candidate fixed-effects cut is too thin to adjudicate.

\begin{table}[tb]
\centering
\caption{Consequences: election-mean divergence does not predict observable governance outcomes (margin, size, year FE; $N=1{,}902$).}
\label{tab:cons}
\small
\begin{tabular}{@{}lcc@{}}
\toprule
Governance outcome & Divergence ($z$) & $p$ \\
\midrule
Activity-and-survival index & $-0.011$ & $0.51$ \\
Log action volume         & $-0.039$ & $0.64$ \\
Action-type breadth       & $+0.008$ & $0.57$ \\
Removed for cause         & $+0.011$ & $0.33$ \\
Tenure (days)             & $+173$   & $0.02$ \\
Still active years later  & $+0.025$ & $0.03$ \\
\bottomrule
\end{tabular}
\\[4pt]
\begin{minipage}{0.62\textwidth}\raggedright\footnotesize Positive longevity rows run opposite to the corruption prediction and read as descriptive (candidate popularity), not causal benefit.\end{minipage}
\end{table}

\begin{figure}[tb]
\centering
\includegraphics[width=0.85\linewidth]{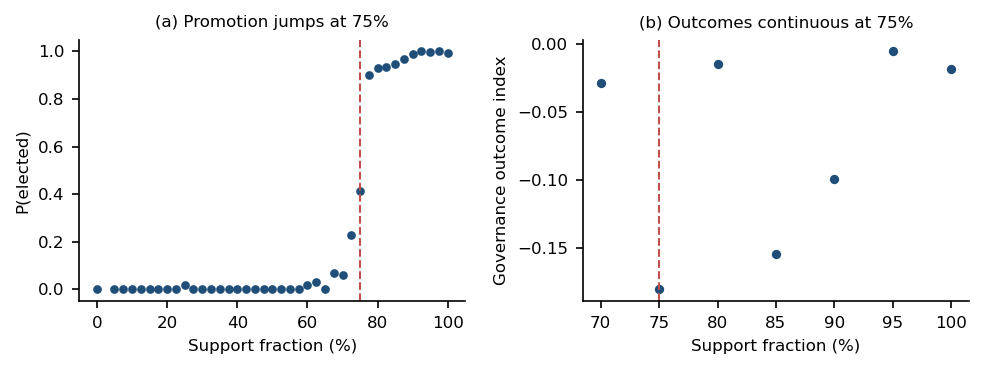}
\caption{Fuzzy RD at the 75\% threshold. Promotion jumps sharply (a); the post-promotion outcome index is continuous (b).}
\label{fig:rd}
\end{figure}

The fuzzy RD agrees for the marginal admit, as Figure~\ref{fig:rd} shows. Panel (a) confirms a sharp first stage, with pass rates jumping from 17.8\% to 81.1\% across the 75\% threshold (instrument coefficient $+0.88$, $p<0.001$, density balanced). Panel (b) shows the governance index is continuous across the same threshold, with no jump at the cutoff. The instrumented effect of being elected on the index is statistically zero across bandwidths, ranging from $-0.38$ to $-0.23$, all with $p>0.13$. The estimates are insignificant, tightly clustered near zero, and stable across bandwidths, so the marginal-admit comparison reads as a bounded null rather than an underpowered one.

Finally, the institution does not correct the signal on its own. Table~\ref{tab:bureau} reports whether bureaucrats discount divergent elections when exercising discretion in the grey zone, where outcomes are not mechanical. They do not. The divergence coefficient on promotion is small and insignificant in every band and under both linear and logistic models. The grey-zone sample is powered to detect a change of about ten percentage points in promotion probability per standard deviation of divergence, and the estimate is near zero. A margin inflated by deferential votes is still read as a genuine mandate, so the burden of correction falls on how the record is displayed and aggregated.

\begin{table}[bp]
\centering
\caption{Bureaucrats do not discount divergence in the grey zone (divergence coef on promotion; controls for support, size, year FE).}
\label{tab:bureau}
\small
\begin{tabular}{@{}lcccc@{}}
\toprule
Support band & $N$ & Promo. & OLS $b$ & $p$ \\
\midrule
65--75\% & 241 & 0.10 & $-0.026$ & 0.46 \\
65--80\% & 336 & 0.30 & $+0.022$ & 0.53 \\
70--80\% & 202 & 0.48 & $+0.032$ & 0.52 \\
\bottomrule
\end{tabular}
\\[4pt]
\begin{minipage}{0.62\textwidth}\raggedright\footnotesize Logistic models agree (all $p>0.36$). Minimum detectable effect at 80\% power is about 0.10 promotion probability per SD of divergence.\end{minipage}
\end{table}

\subsection{Alternative explanations and construct validity}

Several alternatives deserve attention because the construct is rule-based and arrival order is observational. First, late voters might simply be different people. The voter and two-way fixed effects absorb stable voter-level tendencies, so composition does not drive the gradient. Second, the behavior might be cognitive fatigue rather than deference. Fatigue and saturation are closely related, and the prior-text result is consistent with both, but fatigue alone does not predict the substitution toward deference language and away from evidence that the decomposition shows. Third, everything may already have been said, so later voters cite less because earlier voters documented it. This is the form efficient deference takes, and it is the hardest alternative to exclude. It is also where saturation and a rational reading of conformity converge: deferring because the evidence is already on the record is observationally close to herding on that accumulated evidence, and our sign tests, which rule out conformity to the visible count, do not separate the two. We cannot fully separate efficient from empty deference with text alone. We do note that even efficient deference does not restore the margin as a measure of independent scrutiny, since a count of eighty supports in which most defer still rests on a handful of independent assessments. That separation is exactly what the released human-coding instrument is for, and it is the paper's main remaining validity question.

\subsection{Robustness}

Both composites give the same gradient and locus. The gradient survives election, voter, and two-way fixed effects, within-election permutation inference, dropping the length control, and dropping the valence dimension. It also holds inside lopsided elections. Among support votes in races that ended above ninety percent support, divergence still rises with arrival position ($+0.31$, $p<0.001$), so even a near-unanimous margin is built partly from deferential late votes. The consequences null is stable to the alternative composite, to each component separately, and to removing any single year.

\section{Discussion}

\subsection{Theoretical implications}

The result reframes what a governance margin measures. A tally counts votes, but votes carry unequal information, and the inequality is ordered. Later contributors to a public election tend to defer, and the decomposition shows that they trade evidence for agreement language. The pattern points to information saturation rather than to conformity to the visible majority: decisiveness does not amplify the gradient, and accumulated prior text absorbs it. The saturation mechanism connects this finding to IS work on bounded rationality and cognitive load in digital collaboration, where the cost of processing an accumulating record shapes what people contribute \cite{simon1955behavioral,jones2004overload}. The contribution is to separate two layers that the count conflates. The margin can overstate the scrutiny behind it, which is a signal problem. The failure has a precise analogue in composed computational systems, where a chain of individually permitted operations can produce an effect that no single permission authorizes, and the safeguard is a composition rule under which authority may be preserved or lost but never gained \cite{jiang2026chaincaps}. A tally violates that rule: each endorsement is individually legitimate, yet their accumulation confers a mandate that the evidence behind any one of them does not support. The selection itself shows no detectable deterioration in observable outcomes, because the deference is concentrated among peripheral voters who have weaker incentives to invest in costly scrutiny.

The behavior also appears not to reach the decision, at least on the outcomes we can observe. The null on the activity-and-survival index, the bounded null at the RD threshold, and the bureaucrats' non-discounting are together consistent with deference that does not change who governs. We state this as a bound rather than a fact, because our outcomes capture observable activity and survival and cannot rule out unobserved-quality channels; the claim holds for the post-promotion behavior we can measure and is limited to it. We also do not claim a clean causal effect of seeing the tally: the design absorbs election-level and voter-level confounds but not the ordering of behavior itself, so the evidence is consistent with deference produced by sequential public aggregation.

\subsection{Design implications}

The lesson for platform design is bounded but precise. Any system that governs by visible, sequential aggregation, from moderator and maintainer elections to peer-review approvals and up-vote-driven promotion, should not read a large margin as a strong mandate, because the margin can carry less independent scrutiny than its size implies. The right design response follows from the mechanism. Because the driver is saturation of the evidentiary record rather than the decisiveness of the count, the most direct fix is to separate independent evidence from mere endorsement. A platform can let voters mark whether they are adding new evidence or endorsing an existing rationale, summarize the distinct evidentiary claims, and display the number of voters who contribute new evidence alongside the raw support share. Retrieval over a graph-structured record is a natural implementation of this fix, because it represents distinct claims and the relations among them rather than a flat block of text \cite{chen2026survey}; the design goal is to surface which evidentiary claims are new, not to produce one more consensus summary. Collapsing repeated endorsements visually would keep a run of agreement from mimicking independent scrutiny. Hiding or coarsening the running count remains a secondary option. It works against the visibility force, but it does not by itself address saturation, because the discussion record keeps accumulating even when the tally is hidden. The stronger form of that intervention is to preserve independence before aggregation rather than to obscure it afterward: let evaluators commit a rationale in isolation, and surface their agreements and disagreements only once each has done so. Architectures of this kind have been proposed for human--AI research settings, in which agents holding different perspectives investigate independently under a shared provenance contract and their conflicts are adjudicated post hoc by a human decision-maker \cite{zhu2026fundapod}. The same logic applies to a public election: what a margin ought to aggregate is independent evaluations, not evaluations conditioned on one another.

Two further points matter. First, the institution's own gatekeepers do not discount divergence, so any downstream reuse of the margin as a scrutiny signal, whether for confidence, seniority, or legitimacy, carries the distortion forward, and tracking vote--text divergence offers a health metric that participation counts conceal. Second, the problem is forward-looking. If platforms place automated, generative summaries at the top of a thread, the cost to a later participant of aligning with an existing consensus falls further, which our mechanism predicts would deepen saturation. This is the human--AI hybrid setting that IS work on AI governance highlights: when an AI artifact pre-digests the discussion, its summary shapes the human evaluations that feed back into the aggregate \cite{berente2021managing,rai2019hybrids}. Recent work on LLM agents maps this design space as a spectrum running from autonomous operation to human-in-the-loop interaction \cite{su2026agentic}, and the distinction matters here: a summary a participant can interrogate still invites independent scrutiny, whereas a finished, autonomously produced verdict supplies the rationale outright. For example, a generative summary that opens with ``the community broadly supports this candidate'' hands a late voter a ready-made rationale, lowering the cost of writing ``per above'' still further. Deference of this kind also carries a security cost: an artifact that pre-digests the record becomes a single point at which manipulated or injected context can enter the aggregate unchallenged, which is the runtime attack surface that work on agentic systems documents \cite{jiang2026agentic}. Our pre-AI baseline is useful precisely because it measures the human dynamic before that amplification \cite{duan2009informational}.

\section{Conclusion}

Communities that govern by public, sequential voting treat a decisive margin as a strong mandate. We have shown that the margin counts votes while leaving the scrutiny behind them uncounted, and that the gap is ordered. Later votes defer rather than report. The pattern holds under demanding voter and election fixed effects, it tracks the accumulation of the public discussion rather than the decisiveness of the tally, and it is concentrated among network-peripheral voters. The margin is weakened in what it measures, the independent scrutiny behind a choice, while the choice it produces holds up, at least on the outcomes we can observe. For platforms, that turns a design question, how a sequential record is displayed and aggregated, into a question about what their consensus metrics actually measure.

\bibliographystyle{unsrtnat}
\bibliography{refs}

\end{document}